\documentstyle[prl,aps,epsf,floats]{revtex}

\begin{document}


\twocolumn[\hsize\textwidth\columnwidth\hsize\csname
@twocolumnfalse\endcsname

\narrowtext

\begin{bf}Comment on ``Superconducting $\rm PrBa_2Cu_3O_x$''
\end{bf}

\vskip0pc]

Recently, Zou $\em et~al.$ \cite{Zou} reported the observation of 
bulk superconductivity (SC) for a $\rm PrBa_2Cu_3O_x$ (Pr123) single 
crystal grown by the traveling-solvent floating zone (TSFZ) method.  
The SC of Pr123 itself and also the increase of the $T_c$ from 85~K 
to $\approx$~105~K under pressure are of general interest.  These 
unexpected results (see also \cite{Blackstead}) are in sharp contrast 
with the generally accepted view that Pr123 is the only 
nonsuperconducting compound among the orthorhombic $R\rm 
Ba_2Cu_3O_{7-y}$ ($R=$Y, rare earth) cuprates. More detailed
knowledge of TSFZ crystal properties is required to resolve this 
discrepancy.

Zou $\em et~al.$ reported only slight differences in
the crystal structure between TSFZ Pr123 and crystals grown by the 
flux method: an elongation of the $c$-axis parameter 
connected with the expansion of the distance between two $\rm 
CuO_2$ planes was found.  We agree with Zou $\em et~al.$  
that it is hard to attribute the hole delocalization and the 
occurence of SC in the TSFZ crystal to the small elongation of the 
Pr--O(2) distance. Due to the ``strong sample inhomogeneity'' the 
substitution of Ba for Pr might create mobile superconducting holes 
\cite{Zou}.  However, the effective magnetic moment of Pr $\mu_{eff}$ 
was found to be 2.92~$\mu_B$ \cite{Zou}, i.e., close to that of their 
flux crystal.  Thus, given this value of $\mu_{eff}$ it is difficult 
to imagine a substantial substitution of nonmagnetic Ba for Pr.

The aim of this Comment is to show the {\em inconsistency} of the 
value of $\mu_{eff}$ reported by Zou $\em et~al.$ with
their magnetic susceptibility $\chi(T)$ data (Fig.~3 of Ref.\
\cite{Zou}).  For clarity the data of Ref.\ \cite{Zou} 
are shown here together with recent results of our group for a high 
quality Al-free Pr123 single crystal grown in a Pt crucible 
\cite{Narozhnyi}. For this Pr123 crystal the $\chi^{-1}(T)$ curves
are shown in Fig.~1 for the field parallel ($H||c$) and perpendicular 
($H||ab$) to the $c$-axis.  (The field direction for the 
TSFZ crystal was not mentioned in \cite{Zou}).  The values of
$\mu_{eff}$=2.9~$\mu_B$ and 3.1~$\mu_B$ were obtained for our crystal 
for $H||ab$-plane and $H||c$-axis, respectively, from the best fits 
of points at 50~K $\leq T \leq$ 300~K to the modified Curie-Weiss law 
including a temperature independent $\chi_0$ (shown on Fig.~1 by 
solid lines).  These values are in good agreement with previously 
reported ones, see, e.g.  \cite{Hilscher}.  Since it is
impossible to have very closely similar values of $\mu_{eff}$ from 
quite different ``flux'' and ``TSFZ''curves, we reestimated the 
value of $\mu_{eff}$ from the TSFZ data shown in Fig.~1.
 
The first estimate from the $\em linear$ approximation by Zou $\em 
et~al.$ to the $\chi^{-1}(T)$ data (dotted line in Fig.~1 here) gives 
$\mu_{eff}$=2.32~$\mu_B$. According to \cite{Zou} their fit was 
obtained with $\chi_0 \rm=4.5 \times 10^{-4}$~emu/mol. But the $\em 
straight$ line in Fig.~3 of \cite{Zou} representing the fit {\em can~ 
not} be reproduced with that $\chi_0$-value.  The second estimate 
made directly from their data points fitted to the modified 
Curie-Weiss law including $\chi_0$ gives even a smaller value of 
$\mu_{eff}$=2.09~$\mu_B$ and a Curie constant $C$=0.546~emuK/mol. The 
$C$ value for the TSFZ crystal is about one half of that for our flux 
crystal (1.04~emuK/mol and 1.19~emuK/mol for $H||ab$ and $H||c$, 
respectively). This suggests that Pr occupies only about a half of 
the $R$ sites (assuming for the TSFZ crystal nearly the same Pr local 
moment as for the flux grown one).  The other half of the $R$ sites 
is occupied most probably by the nonmagnetic Ba. Noteworthy, SC with 
$T_c \approx$~43~K was observed for $\rm 
Pr_{0.5}Ca_{0.5}Ba_2Cu_3O_{7-y}$ thin films \cite{Norton}.  Ba$^{2+}$ 
as well as Ca$^{2+}$ on $R$-site dopes additional mobile holes and 
compensates for the localization of holes by the Pr--O(2,3) 
hybridization.  Ba$^{2+}$ has a larger ionic radius than Pr$^{3+}$ 
and so the substitution of Ba for Pr could give a natural explanation 
not only for the SC in TSFZ Pr-123 but also for the elongation of the 
distance between the CuO$_2$ planes observed in \cite{Zou}.

After the completion of the present Comment we have learnt, that the 
possibility of Ba substitution on Pr-site was also pointed out in 
\cite{Pieper_cm} in the context of NMR-data.

\begin{figure} 
\epsfysize=6.5cm 
\centerline{\epsfbox{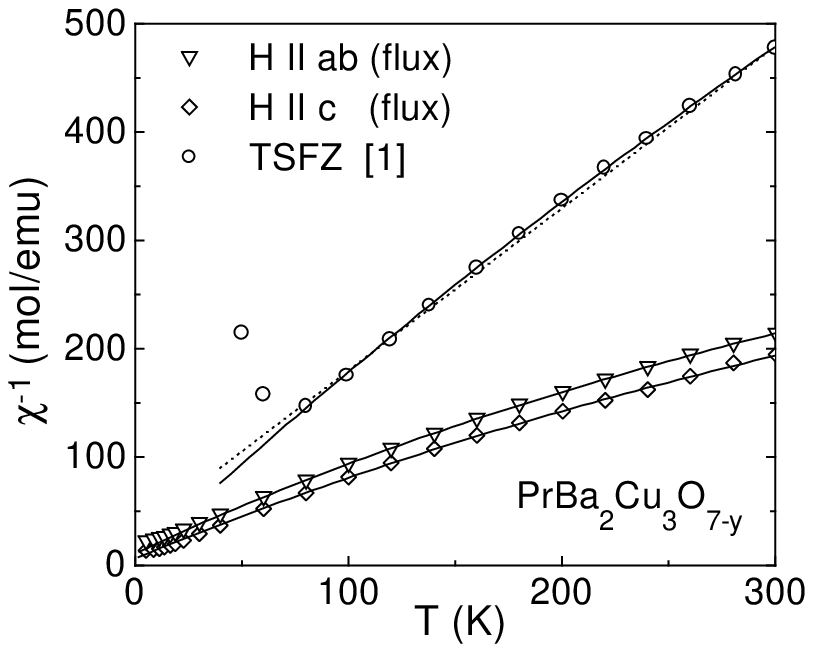}}
\caption{$\chi^{-1}$ vs $T$ for flux grown \protect \cite{Narozhnyi} 
and TSFZ \protect \cite{Zou} Pr-123 single crystals. Solid lines - 
fits to the Curie-Weiss law. Only some representative 
points are shown. For details see text .} \label{fig1}
\end{figure}

We thank G. Fuchs and K.-H. M\"uller for useful discussions.
This work was supported by RFBR grants 96-02-00046G, 96-02-16305a; 
DFG grant MU1015/4-1.

 
V. N. Narozhnyi$^{1,2,\ast}$ and S.-L. Drechsler$^2$  

$^1$Institute for High Pressure Physics, Russian Academy of 
Sciences, Troitsk, Moscow Region, 142092, Russia, 

$^2$Institut f\"ur Festk\"orper- und Werkstofforschung 
Dresden e.V., Postfach 270016, D-01171 Dresden, Germany
 
 

\end{document}